\documentclass[twocolumn,showpacs,preprintnumbers]{revtex4}
\usepackage{amssymb}
\usepackage{amsfonts}
\usepackage{amsmath}
\usepackage{graphicx}
\usepackage{dcolumn}
\usepackage{bm}

\input{tcilatex}
\begin{document}

\preprint{APS/123-QED}
\title{Full elimination of the gravity-gradient terms in atom interferometry}
\author{B. Dubetsky}
\email{bdubetsky@gmail.com}
\affiliation{1849 S Ocean Dr, Apt 207, Hallandale, FL 33009}
\date{\today }

\begin{abstract}
The A. Roura technique was modified to eliminate all terms in the atom
interferometer phase, which are linear in the gravity-gradient tensor. The
full elimination occurs if all effective wave vectors are slightly changed.
The full elimination technique would allow to relieve the synchronization
requirements in the test of the Einstein equivalence principle. This
technique also eliminates the error of the absolute gravity measurement
associated with the gravity gradient terms. The error becomes three orders
smaller and does not depend on the time delay between the Raman pulses. In
addition, the new differential scheme is proposed to observe the
gravity-gradient term independent on the atoms initial position and velocity.
\end{abstract}

\maketitle

\section{Introduction.}

Atom interferometers \cite{c0} are used as sensors of the gravity field. In
the Earth gravity field $\mathbf{g},$ the main contribution to the atom
interferometer phase is given by \cite{c0.1}%
\begin{equation}
\phi =\mathbf{k}\cdot \mathbf{g}T^{2},  \label{1}
\end{equation}%
where $\mathbf{k}$ is the effective wave vector of the Raman pulses used as
the beam splitters, $T$ is the time separation between pulses. This phase is
used to measure the local gravity field \cite{c1,c3,c4}, the gravity
gradient tensor \cite{c4.1}, the Newtonian gravitational constant \cite%
{c4.2,c4.3}, to test the Einstein's Equivalence Principle (EEP) \cite%
{c4.4,c4.5,c4.6}. For the local gravity measurement, one has to get the
absolute value of the phase, while for the other applications one can use
the differential schemes. If the field is slightly non-uniform then the
phase becomes sensitive to the atomic initial position $\mathbf{x}$ \cite%
{c1.0} and velocity $\mathbf{v}$ \cite{c6,c5}. Using large interrogation
time $T$ one can increase the precision of the measurements, but
simultaneously the gravity field slightly changes along the atomic
trajectories, which leads to the one more contribution, to the phase $\phi
_{WT},$ which was first found by P. Wolf and P. Tourrence \cite{c5.1}.
Keeping together all contributions and allowing time delay $T_{1}$ between
the moment of atom clouds launching and the first Raman pulse one finds

\begin{subequations}
\label{2}
\begin{eqnarray}
\phi &=&\mathbf{k}\cdot \mathbf{g}T^{2}+\phi _{R}+\phi _{WT},  \label{2a} \\
\phi _{R} &=&\mathbf{k}\cdot \Gamma T^{2}\left( \mathbf{x}+\mathbf{v}%
T\right) ,  \label{2b} \\
\phi _{WT} &=&\mathbf{k}\cdot \Gamma \mathbf{g}T^{2}\left( \dfrac{7}{12}%
T^{2}+TT_{1}+\dfrac{1}{2}T_{1}^{2}\right) ,  \label{2c}
\end{eqnarray}%
where $\Gamma $ is the gravity gradient tensor. In this expression we
include only zero-order and linear terms in $\Gamma ,$ did not take into
account the Earth rotation and quantum corrections. To eliminate the
sensitivity to the initial position and velocity, term $\phi _{R}$, A. Roura 
\cite{c1.0} proposed to assign the effective wave vector of the second pulse
to be slightly different from $\mathbf{k},$%
\end{subequations}
\begin{equation}
\mathbf{k}_{2}=\left( 1+\Gamma T^{2}/2\right) \mathbf{k}.  \label{3}
\end{equation}%
This effect has been recently observed \cite{c7,c8}.

\section{Non-elliminated term}

I noted \cite{c1.1} that the WT-term $\phi _{WT}$ is not eliminated.\ After
using A. Roura technique, one finds \cite{c1.1}%
\begin{equation}
\phi _{WT}=\dfrac{1}{12}\mathbf{k}\cdot \Gamma \mathbf{g}T^{4},  \label{4}
\end{equation}%
which means that dependence on the time $T_{1}$ is eliminated, but
dependence on the time $T$ is eliminated only partially, becomes seven times
smaller. Analysis shows that for $T_{1}=0$ this partial elimination is
caused by the difference between an atomic acceleration extracted from the
expression for the interferometer phase and the actual atomic acceleration
at the middle time $t=T.$ The difference was first studied in the article 
\cite{c6.1}. WT-term leads to the errors in the absolute gravimetry and in
the EEP test. Let us consider these errors separately.

\subsection{Absolute gravimeter systematic error}

For the precision interferometry one should increase the time $T.$ The time
as large as $T=1.15s$ was achieved in the article \cite{c1.2}. For $\Gamma
\approx 3\cdot 10^{-6}$s$^{-2}$, $k=1.6\cdot 10^{7}$m$^{-1},T_{1}=0,$ $\phi
_{WT}\approx 70$rad, which would affect the absolute gravity measurement on
the level $0.3$ppm.

To suppress the influence of the term (\ref{4}) on the gravity measurement,
it was proposed \cite{c1.3} to define\ $g$ as%
\begin{equation}
g=\phi \left\{ kT^{2}\left[ 1+\left( 1/12\right) \Gamma _{zz}T^{2}\right]
\right\} ^{-1},  \label{4.1}
\end{equation}%
where $z$ is the vertical axis and $\mathbf{k}\Vert \mathbf{g}$. But,
evidently, in this case error will be caused by the uncertainty of the
gravity-gradient measurement $\delta \Gamma _{zz},$%
\begin{equation}
\delta g\thicksim \dfrac{1}{12}\delta \Gamma _{zz}T^{2}g.  \label{4.2}
\end{equation}%
Though, starting from the article \cite{c4.1}, there are a lot of papers
devoted to the atom interferometry based gradiometry, I know only two
articles \cite{c1.4,c1.5}, where the value of $\Gamma _{zz}$ has been
published. The uncertainty $\delta \Gamma _{zz}=30E$ has been achieved \cite%
{c1.5}. This uncertainty leads to the error $\delta g\thicksim 3$ng. Since
WT-term increases as $T^{4},$ the relative weight of the error (\ref{4.2})
also increases as $T^{2}$ for the lager precision.

\subsection{EEP-test}

The partial elimination, the presence of the term (\ref{4}), could result
also in the restriction of some importance for the test of the EEP. Since
one uses two atomic species, $A$ and $B,$ which could have different
effective wave vectors $\mathbf{k}_{A}$ and $\mathbf{k}_{B}$ one has to have
different Raman pulses, which could be asynchronized, their time delays
between pulses $T_{A}$ and $T_{B}$ could be slightly different. If both wave
vectors are vertical and one measures the parameter%
\begin{equation}
\eta ^{\prime }=\dfrac{2\left( \dfrac{\phi _{A}}{k_{A}}-\dfrac{\phi _{B}}{%
k_{B}}\right) }{\dfrac{\phi _{A}}{k_{A}}+\dfrac{\phi _{B}}{k_{B}}}
\label{4.3}
\end{equation}%
Even neglecting WT-term, one finds from Eq. (\ref{1}) that%
\begin{equation}
\eta ^{\prime }\approx \eta +\dfrac{T_{A}^{2}-T_{B}^{2}}{T_{A}^{2}+T_{B}^{2}}
\label{4.4}
\end{equation}%
where%
\begin{equation}
\eta =\dfrac{2\left( g_{A}-g_{B}\right) }{g_{A}+g_{B}}  \label{6.1}
\end{equation}%
is the E\"{o}tv\"{o}s-parameter, and we neglect the second order terms
proportional to $\left( g_{A}-g_{B}\right) \left( T_{A}^{2}-T_{B}^{2}\right)
.$ For the current state of art in the atomic interferometry, one should be
able to measure the parameter $\eta ^{\prime }$ with inaccuracy $\delta \eta
=1.5\cdot 10^{-14}$ at $T=1.04$s \cite{c9}. One sees that for a proper use
of the parameter (\ref{4.3}) to test EEP one has to synchronize Raman pulses
with accuracy%
\begin{equation}
\left\vert T_{A}-T_{B}\right\vert \lesssim \delta \eta T  \label{4.5}
\end{equation}%
To avoid this severe restriction one can measure the parameter%
\begin{equation}
\eta ^{\prime \prime }=\dfrac{2\left( \dfrac{\phi _{A}}{k_{A}T_{A}^{2}}-%
\dfrac{\phi _{B}}{k_{B}T_{B}^{2}}\right) }{\dfrac{\phi _{A}}{k_{A}T_{A}^{2}}+%
\dfrac{\phi _{B}}{k_{B}T_{B}^{2}}},  \label{5}
\end{equation}%
then, owing to the WT-term (\ref{4}) one gets for this parameter%
\begin{equation}
\eta ^{\prime \prime }\approx \eta \left[ 1+\dfrac{1}{24}\Gamma _{zz}\left(
T_{A}^{2}+T_{B}^{2}\right) \right] +\dfrac{1}{12}\Gamma _{zz}\left(
T_{A}+T_{B}\right) \left( T_{A}-T_{B}\right) ,  \label{6}
\end{equation}%
One sees that to avoid the systematic error caused by the term (\ref{4}) one
has to synchronize Raman pulses better than 
\begin{equation}
\left\vert T_{A}-T_{B}\right\vert \lesssim \dfrac{6\delta \eta }{\Gamma
_{zz}T}\approx 30\text{ns}  \label{6.2}
\end{equation}%
This time is 3 order of magnitude less than the typical Raman pulses'
duration.

\section{Full elimination}

To avoid the error (\ref{4.2}) in the absolute gravity measurements and the
restriction (\ref{6.2}), it should be useful to eliminate all terms (\ref{2b}%
, \ref{2c}). For this purpose, I propose to extend the A. Roura technique 
\cite{c1.0} and to change the effective wave vectors of all 3 Raman pulses, 
\begin{equation}
\mathbf{k}_{i}=\mathbf{k}+\Delta \mathbf{k}_{i},  \label{7}
\end{equation}%
where $\Delta \mathbf{k}_{i}$ are linear in the tensor $\Gamma .$ Using
expression for the phase of the Mach-Zehnder atom interferometer with
different wave vectors of the Raman pulses \cite{c2}, in the limit $\hbar
\rightarrow 0,$%
\begin{equation}
\phi =\mathbf{k}_{1}\cdot \mathbf{x}\left( T_{1}\right) -2\mathbf{k}%
_{2}\cdot \mathbf{x}\left( T_{1}+T\right) +\mathbf{k}_{3}\cdot \mathbf{x}%
\left( T_{1}+2T\right) ,  \label{8}
\end{equation}%
where 
\begin{equation}
\mathbf{x}\left( t\right) =\mathbf{x}+\mathbf{v}t+\dfrac{1}{2}\mathbf{g}%
t^{2}+\Gamma \left[ \dfrac{1}{2}\mathbf{x}t^{2}+\dfrac{1}{6}\mathbf{v}t^{3}+%
\dfrac{1}{24}\mathbf{g}t^{4}\right] +\ldots   \label{9}
\end{equation}%
is an atom trajectory, in which I hold only the contributions to the lower
order in $\Gamma $ \cite{c10}, and requiring that all terms proportional to $%
\mathbf{x},~\mathbf{v}$ and $\Gamma \mathbf{g}$ have to be eliminated, one
gets 3 linear equations for the wave vectors' changes $\Delta \mathbf{k}_{i}.
$ The solution of those equations is 
\begin{subequations}
\label{10}
\begin{eqnarray}
\Delta \mathbf{k}_{1} &=&\Delta \mathbf{k}_{3}=-\dfrac{1}{12}\Gamma T^{2}%
\mathbf{k},  \label{10a} \\
\Delta \mathbf{k}_{2} &=&\dfrac{5}{12}\Gamma T^{2}\mathbf{k}.  \label{10b}
\end{eqnarray}%
We verified that this choice of the wave vectors will also eliminate the
quantum contribution to the phase \cite{c6,c5} of the order of $\dfrac{\hbar 
}{M}\mathbf{k\cdot }\Gamma \mathbf{k}T^{3},$ where $M$ is an atomic mass.

\section{Error analysis}

Let us estimate now the error of the full elimination technique for the
absolute gravimetry. Ideally, the changes of the wave vectors (\ref{10})
make gravity measurements insensitive to the linear in tensor $\Gamma $
terms. But, since $\Gamma $ is known with a limited inaccuracy, the
elimination (\ref{10}) leads also to some error in the phase. If one changes
the wave vectors as $\Delta \mathbf{k}_{1}=\Delta \mathbf{k}_{3}=-\mathbf{%
\alpha ,}$ $\Delta \mathbf{k}_{2}=5\mathbf{\alpha }$, where $\mathbf{\alpha }
$ is a vector fitting parameter, which varies near the desired value 
\end{subequations}
\begin{equation}
\mathbf{\alpha =}\dfrac{1}{12}\Gamma T^{2}\mathbf{k+}\delta \mathbf{\alpha ,}
\label{11}
\end{equation}%
then one finds for the phase (\ref{8})%
\begin{eqnarray}
\phi  &=&\mathbf{k\cdot g}T^{2}-12\delta \mathbf{\alpha \cdot x}-12\left(
T_{1}+T\right) \delta \mathbf{\alpha \cdot }\dfrac{\mathbf{p}}{M}  \notag \\
&&-\left( 7T^{2}+12T_{1}T+6T_{1}^{2}\right) \delta \mathbf{\alpha \cdot g}.
\label{11.1}
\end{eqnarray}%
One can initially use the differential technique \cite{c7} to find $\mathbf{%
\alpha .}$ The phase difference of the two identical interferometers with
atom clouds launched from the points $\mathbf{x}_{1}$ and $\mathbf{x}_{2}$
is used here. If the distance between clouds $\mathbf{L=x}_{1}-\mathbf{x}_{2}
$, wave vector and launching momenta are directed along the $z-$axis, then
one gets%
\begin{equation}
\delta \alpha _{z}=-\dfrac{\delta \phi }{12L},  \label{11.2}
\end{equation}%
where $\delta \phi $ is the phase difference, which could be as small as the
phase noise. For the phase noise $\delta \phi \thicksim 10^{-3},$ the
distance $L\thicksim 10$m, the uncertainty of the fitting parameter $\delta
\alpha _{z}\thicksim 10^{-5}$m$^{-1}.$ Then for the gravity $g=\phi /kT^{2}$
one obtains from Eqs. (\ref{11.1}, \ref{11.2}) that the error is given by%
\begin{equation}
\delta g\approx \dfrac{5\delta \phi }{12kL}g\thicksim 3\text{pg},
\label{11.3}
\end{equation}%
where we assumed that $x\ll gT^{2},~T\gg T_{1}$ and one exploits fountain
technique \cite{c11}, $\mathbf{p}=-M\mathbf{g}T.$ Comparing this error with
the error (\ref{4.2}) of the technique proposed in \cite{c1.3}, one sees
that using our proposal of the full elimination of the gravity-gradient
terms decreases the error on the $3$ orders of magnitude. Moreover, since
the error (\ref{11.3}) is $T-$independent, it stays the same for the higher
precision.

\section{New differential scheme to observe the WT-term.}

Let us return \ to the WT-term (\ref{2c}). Being independent on the atomic
initial position and velocity, the term could not be observed in the
previously used differential schemes \cite{c4.1,c4.2,c4.3}. But if in the
same differential scheme atom clouds are launched at the slightly different
moments, the time $T_{1}$ before the first Raman pulse for the first cloud
and the time $\delta T_{1}$ earlier for the second cloud and if time $\delta
T_{1}$ is sufficiently small to neglect the vibrational noise during that
time, then one gets following term in the phase difference

\begin{equation}
\delta \phi =\mathbf{k}\cdot \Gamma \mathbf{g}T^{2}\left( T+T_{1}+\delta
T_{1}/2\right) \delta T_{1}.  \label{12}
\end{equation}%
The term (\ref{12}) can be canceled using any elimination technique.

\section{Conclusion}

In a conclusion, we propose to modify the A. Roura technique \cite{c1.0},
i.e. to change the wave vectors of all the Raman pulses for the elimination
of all the terms linear in the gravity-gradient tensor. This full
elimination technique allows one to increase the accuracy of the absolute
gravity measurements and to avoid the synchronization of the Raman pulses
used to test EEP. In addition, we propose to modify the differential scheme
to observe in the phase difference the term independent on the differences
of the atom clouds' initial position and velocity.

\end{document}